\documentclass[a4paper]{JACoW-GSI}
\usepackage{graphicx}
\usepackage{url}

\begin{document}
\title{CP-tagged charm decays: relevance, status and prospects}

\author[1]{G. Wilkinson}
\affil[1]{University of Oxford}

\maketitle

\abstract{The analysis of quantum-correlated $D-\bar{D}$ 
decays produced at the $\psi(3770)$ resonance gives
unique insight into quantities such as strong-phase
differences and coherence factors.  Knowledge of these
parameters is invaluable for measurements of the
CKM-angle $\gamma$ ($\phi_3$) in $B \to DK$ decays.
Results from CLEO-c analyses performed at the $\psi(3770)$ 
resonance in a variety of decay channels
are reported, and their consequences for the determination
of $\gamma$ is assessed.  Future prospects are given for
extensions to the present studies.}

\section{Overview}

In the last couple of years results have begun to emerge from the
$\psi(3770)$ dataset of the CLEO-c experiment which exploit 
the quantum-correlated nature of the $D-\bar{D}$ production at
this resonance.   These results provide insight into the strong-phase
differences existing between $D^0$ and $\bar{D^0}$ decays which
is of great interest for various applications, in particular the
measurement of the CP-violating unitarity triangle angle 
$\gamma$ ($\phi_3$) in $B$-decays.

This review is organised as follows.  First the relevance
of $CP$-tagging is explained, and the basic analysis principle
is presented.  Results, both preliminary and final, are
then shown in two categories of $D$-decay: $D \to K^0 h^+h^-$ ($h = \pi$ or $K$)~\footnote{In this article $D$ will be used to
indicate any neutral $D$-meson.}
and $D \to K n \pi$ ($= K^\pm \pi^\mp, \, K^\pm \pi^\mp \pi^+\pi^-$ and $ K^\pm\pi^\mp \pi^0 $).
Particular attention is paid to the impact these measurements will have
on the determination of $\gamma$. 
Finally, a summary is given, along with future prospects.

\section{Preliminaries}

\subsection{Importance of CP-tagged $D$-decays}

CP-tagged $D$ decays access information which is not available from
decays of flavour-tagged mesons, namely the strong-phase
difference between $D^0$ and $\bar{D^0}$ decay to the
final state of interest.  As a $D$-meson in a CP-eigenstate 
is a superposition of $D^0$ and $\bar{D^0}$, the decay probability 
has a dependence on the cosine of this strong-phase difference.
In a two-body $D$-decay this strong-phase difference is a single quantity,
whereas for three or more particles it will vary  
over Dalitz space, depending on the intermediate resonances contributing
at each position.

Knowledge of strong-phase differences (hereafter `strong phases')
is important for three reasons.
\begin{itemize}
\item{It is interesting in itself for understanding $D$-decay dynamics and the
resulting light-quark mesons produced.}
\item{It is necessary to relate various measurements of the $D$-mixing parameters
$x$ and $y$ (where $x$ characterises the mass splitting between the mass eigenstates,
and $y$ the width splitting).  For example in the `wrong sign' $D \to K\pi$ mixing
analysis the measured parameters differ from $x$ and $y$ through the rotation 
by the strong phase $\delta_D^{K\pi}$.}
\item{It is invaluable for measurements of the CP-violating angle $\gamma$ ($\phi_3$). }
\end{itemize}
It is the last point which provides the main context for the
discussion in the present review.

\subsection{Measuring $\gamma$ in $B \to D K$ decays}

The angle $\gamma$ is the least well-known parameter of the CKM unitarity triangle 
with an uncertainty presently estimated to be around $30^\circ$~\cite{CKMFITTER}.
All measurements contributing to our existing knowledge come from
the $B$-factory experiments, and have been performed using the so-called 
`$B \to DK$' family of methods.  This approach will continue to dominate
the $\gamma$ determination at the LHCb experiment, where the increased
$B$-decay statistics will allow an order of magnitude improvement in precision~\cite{AKIBA}.

A $B^-$ meson may decay to $D^0 K^-$ through a $b \to c$ transition,
or $\bar{D^0} K^-$ through a $b \to u$ transition.  If the charm meson
is reconstructed in any mode common to both $D^0$ and $\bar{D^0}$ 
(examples include $K^0_S \pi^+\pi^-$ and $K^\pm\pi^\mp$) then
interference occurs which involves $\gamma$, the CP-violating phase between
the two $b$-decay paths.  Therefore measuring the difference in
rates (or kinematical distributions in the case of three-or-more-body channels) 
between $B^-$ and $B^+$ decays gives sensitivity to this angle.

In order to extract $\gamma$, however, other parameters must be 
accounted for which also affect the interference.  These 
include $r_B$, the ratio of the magnitude of the $B$-decay amplitudes,
$\delta_B$, the strong-phase difference between the $B$-decay amplitudes,
and $\delta_D$ the strong phase (or phases) associated with the $D$-decay.
Although in some cases it is in principle possible to use 
the $B$-data themselves to extract the $B$- and $D$-decay parameters along with $\gamma$,
it is generally advantageous, and often essential, to have an external constraint (or constraints) 
on $\delta_D$.  The best source of this information is CP-tagged $D$-decays.

\subsection{Quantum-correlated $\psi(3770)$ decays}

The most practical source of CP-tagged $D$ decays
are neutral $D - \bar{D}$ events produced at the $\psi(3770)$ resonance 
in which one meson (the `signal $D$') decays to the final state of interest, and the other is
reconstructed in a CP-eigenstate (the `tagging $D$').  The $D - \bar{D}$ system is produced
in a coherent state which, due to the quantum numbers of the resonance, is known
to be C-odd.  Therefore if one $D$ is reconstructed as CP-even, for example through 
$D \to K^+K^-$, then the other meson is `tagged' as being in a CP-odd state.

Reconstructing both $D$-mesons (`double-tagging') in $\psi(3770)$ decays allows for
studies to be generalised beyond pure CP-tagging.  As all hadronic final states
can be accessed from both $D^0$ and $\bar{D^0}$ decays this means that even
if the tagging $D$ is reconstructed in a non-CP eigenstate hadronic mode then
the signal $D$ will also be in some superposition of $D^0$ and $\bar{D^0}$,
albeit not in the equal proportions that are present in the CP-tagged case.  Such events turn
out to be a powerful addition to the pure CP-tagged sample and are extensively
used in the analyses described in this review.  For this reason the subsequent discussion
will often refer to `quantum-correlated decays'.

The only existing $\psi(3770)$ dataset of significant size  was collected by
the CLEO-c experiment in $e^+e^-$ collisions at 
the Cornell Electron Storage Ring (CESR). CLEO-c finished
operation in Spring 2008, by which time it had accumulated 818~$\rm pb^{-1}$ of 
data at the $\psi(3770)$ resonance, together with additional integrated luminosity
at other centre-of-mass energies.  All analyses presented in this review use
this dataset.   In the near future, higher statistics samples 
are expected from BES-III~\cite{BES}.

There are other important side-benefits to running at this threshold energy of 3770~MeV.  
Events  are very clean, without fragmentation debris.  If all 
$D$-decay charged particles and photons are identified in the event, the kinematical 
constraints allow  the presence of neutral particles such as $K^0_L$ or indeed 
neutrinos to be inferred.  This feature enables the range of CP-tags to be essentially
doubled with respect to those available with normal reconstruction techniques.  For
example $K^0_L \pi^0$ decays may be used as well as $K^0_S \pi^0$.  Furthermore, 
signal decays involving $K^0$ mesons may be studied in both the $K^0_S$ and $K^0_L$ categories.

\section{Quantum correlated studies of $D \to K^0_S h^+ h^-$ decays}

\subsection{Model dependent measurement of $\gamma$ in \\ $B \to D (K^0_S \pi^+\pi^-)K$ decays}

With the statistics presently available at the $B$-factory experiments
the highest sensitivity to the angle $\gamma$ is found in the channel $B \to D (K^0_S \pi^+\pi^-)K$.
Comparison of the $K^0_S \pi^+\pi^-$ Dalitz space in decays originating from a
$B^-$ with those from a $B^+$ meson reveals CP-violating differences which,
when analysed in an unbinned likelihood fit, can be used to extract a value for $\gamma$.
Using this approach BABAR obtain a result 
of $\gamma = (76 \pm 22 \pm 5  \pm 5 )^\circ$~\cite{BABARDALITZ}
and Belle $\gamma = (76^{+12}_{-13} \pm 4 \pm 9)^\circ$~\cite{BELLEDALITZ}.
Here the uncertainties are statistical, systematic and model errors, respectively~\footnote{
It may be noted that there is a significant
difference in the reported statistical precision in the two results which
cannot be explained by the variation in sample sizes between the experiments. 
This difference is largely driven by the very different values of
the interference parameter $r_B$ that
is found in the two analyses.}. In fact rhe BABAR result also receives a
contribution from the analysis of $B \to D (K^0_S K^+K-)K$ events;
the model error for $B \to D (K^0_S \pi^+\pi^-)K$ alone is estimated to be $\sim 7^\circ$.

The model uncertainty in these analyses represents the limit in the understanding
of the $D$-meson decay.  The likelihood function used in the fit includes
a description of the $D^0 \to K^0_S \pi^+\pi^-$ decay which is modelled from
a sample of flavour-tagged $D^{\ast +} \to D^0 (K^0_S \pi^+\pi^-) \pi^+$ events.
Both BABAR and Belle have developed their own models of this decay.
The BABAR model, for example, derives from 487,000 events and is based
on the isobar formalism~\cite{KOPP}, with the S-wave $\pi\pi$ and $K\pi$
contributions treated with the K-matrix~\cite{KMATRIX} and LASS~\cite{LASS}
approaches respectively.  The agreement of data with this model is very impressive,
but inevitably is not perfect.  (The $\chi^2/n.d.f.$  is
found to be 1.11 for $\sim 19k$ degrees of freedom.)   Although a $7-9^\circ$ model
uncertainty is at present adequate in the measurement of $\gamma$ 
given the $B$-meson decay statistics available at the $B$-factories, 
it will rapidly become a limiting factor to 
the precision of the same analysis performed at LHCb, where a few degree statistical
uncertainty is foreseen with $10\,{\rm fb^{-1}}$~\cite{LAZZERONI}.  A model independent
approach is therefore highly desirable.
 
\subsection{Model independent measurement of $\gamma$ in \\ $B \to D (K^0_S \pi^+\pi^-)K$ decays}

A binned Dalitz analysis approach to the $\gamma$ determination in  
$B\to D (K^0_S \pi^+\pi^-)K$~\cite{GIRI,BONDAR} removes any model
dependence by relating the number of events observed in a given bin 
of Dalitz space to {\it experimental observables}.  If the Dalitz plot
is partitioned into a set of bins symmetric through 
the line $m^2(K^0_S \pi^+) = m^2(K^0_S \pi^-)$, then the number of $B^\pm$ events
giving rise to decays in bin $i$ is given by:
\begin{equation}
N_i^\pm = h  \left( K_{i}  +  r_B^2 K_{-i}  +  2  \sqrt{K_i K_{-i}} (x_\pm c_i \pm y_\pm s_i) \right)
\label{eq:dalitzbin}
\end{equation}
where $h$ is a normalisation factor, $x_\pm =r_B \cos (\delta_B \pm \gamma )$, 
$y_\pm =r_B \sin (\delta_B \pm \gamma )$, $K_i$ are the number of flavour-tagged $D^0 \to K^0_S \pi^+\pi^-$
decays in bin $i$, and $c_i$ and $s_i$ are the amplitude averaged cosine and sine of the strong
phase of the $D$-decay in the bin in question.  In this expression the subscript $-i$ indicates a bin which
is defined symmetrically in the lower region of the Dalitz plot with respect to bin $i$ in the upper region. 
The parameters $\gamma$, $r_B$ and $\delta_ B$
are to be extracted from the measurement, while $c_i$ and $s_i$ are inputs
which are determined from quantum-correlated $D$-decays.  The values of $K_i$ can
be determined from any sample of flavour-tagged $D$-decays.

\begin{figure}[htb]
\centering
\includegraphics*[width=0.40\textwidth]{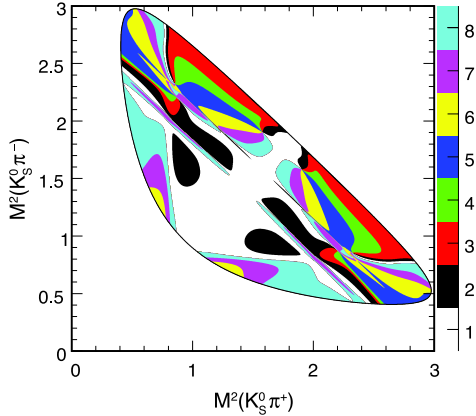}
\caption{Equal phase binning of the $D^0 \to K^0_S \pi^+\pi^-$ Dalitz plot.}
\label{fig:kpipi_binning}
\end{figure}

In making such a measurement the statistical sensitivity is affected 
by the choice of binning.  It is advantageous to define bins in which
the variation of strong phase is small (so that $c_i^2 + s_i^2$ is as
close as possible to $1$).   This choice can be informed by the
models developed on the flavour-tagged data.  It must be emphasised, however,
that any difference between the model and reality will {\it not} lead
to any bias in the measurement, but will merely result in the statistical
sensitivity being lower than expected.    In the analysis reported below
8 bins are chosen, each covering the same span in strong phase.
The model used to make this choice is that constructed by BABAR~\cite{BABARDALITZ}.
This binning is shown in Fig.~\ref{fig:kpipi_binning}.  

\subsection{Measurement of $c_i$ and $s_i$ in $D \to K^0_S \pi^+\pi^-$}

\begin{figure}[htb]
\centering
\includegraphics*[width=0.40\textwidth]{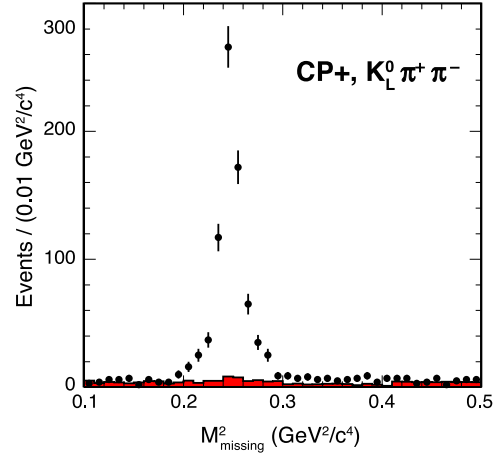}
\caption{CLEO-c missing mass squared distribution in $K^0_L\pi^+\pi^-$ events
reconstructed together with CP-even tags.  The shaded histogram 
represents the background expectation from the simulation.}
\label{fig:klrec}
\end{figure}

The parameters $c_i$ and $s_i$ have been determined using $818~\rm{fb^{-1}}$
of $\psi(3770)$ data collected by CLEO-c~\cite{CLEOCCISI}.  Central to the
analysis is a sample of double-tagged events in which $D \to K^0_S \pi^+\pi^-$
decays are reconstructed together with a CP-eigenstate, or against other
$D \to K^0_S \pi^+\pi^-$ decays. $D \to K^0_L \pi^+\pi^-$ decays are also
used.   The presence of a $K^0_L$ mesons is inferred by
selecting events in which the missing-mass squared is consistent
with $m_{K^0_L}^2$, as illustrated in Fig.~\ref{fig:klrec}

A summary of the double-tag sample is given in Tab.~\ref{tab:kpipiyield}.
Four CP-even tag and three CP-odd tag modes are employed alongside $D \to K^0_S \pi^+\pi^-$ 
decays.  Background considerations mean that certain CP-tags are not used with
the $D \to K^0_L \pi^+\pi^-$ decays.  Approximately 1600 CP-tagged events are selected 
in total, and around 1300 $K^0_S \pi^+\pi^-$ vs $K^0 \pi^+\pi^-$ events.   The signal
to background level is between 10 and 100, depending on the tag mode.
Also selected (but not tabulated here) are events in which the signal
is reconstructed alongside decays such as $D \to K^\pm \pi^\mp$, which to a very
good approximation serve as flavour-tags.

\begin{table}
\caption{CLEO-c double tag yields in the $K^0 \pi^+\pi^-$ analysis.}\label{tab:kpipiyield}
\begin{center}
\begin{tabular}{lcc}\\ \hline \hline
Mode   & $K^0_S \pi^+\pi^-$ yield & $K^0_L \pi^+\pi^-$ yield \\ \hline
\multicolumn{3}{c}{CP-even tags} \\ \hline
$K^+K^-$           & 124 & 345 \\
$\pi^+\pi^-$       &  62 & 172 \\
$K^0_S \pi^0\pi^0$ &  56 &  -  \\
$K^0_L\pi^0$       & 229 &  -  \\ \hline
\multicolumn{3}{c}{CP-odd tags} \\ \hline
$K^0_S \pi^0$      & 189 & 281 \\
$K^0_S \eta$       &  39 &  41 \\
$K^0_S \omega$     &  83 &   - \\ \hline
\multicolumn{3}{c}{Other tags} \\ \hline
$K^0_S \pi^+\pi^-$ & 475 &  867 \\ \hline \hline
\end{tabular}
\end{center}
\end{table}

\begin{figure}[htb]
\centering
\includegraphics*[width=0.48\textwidth]{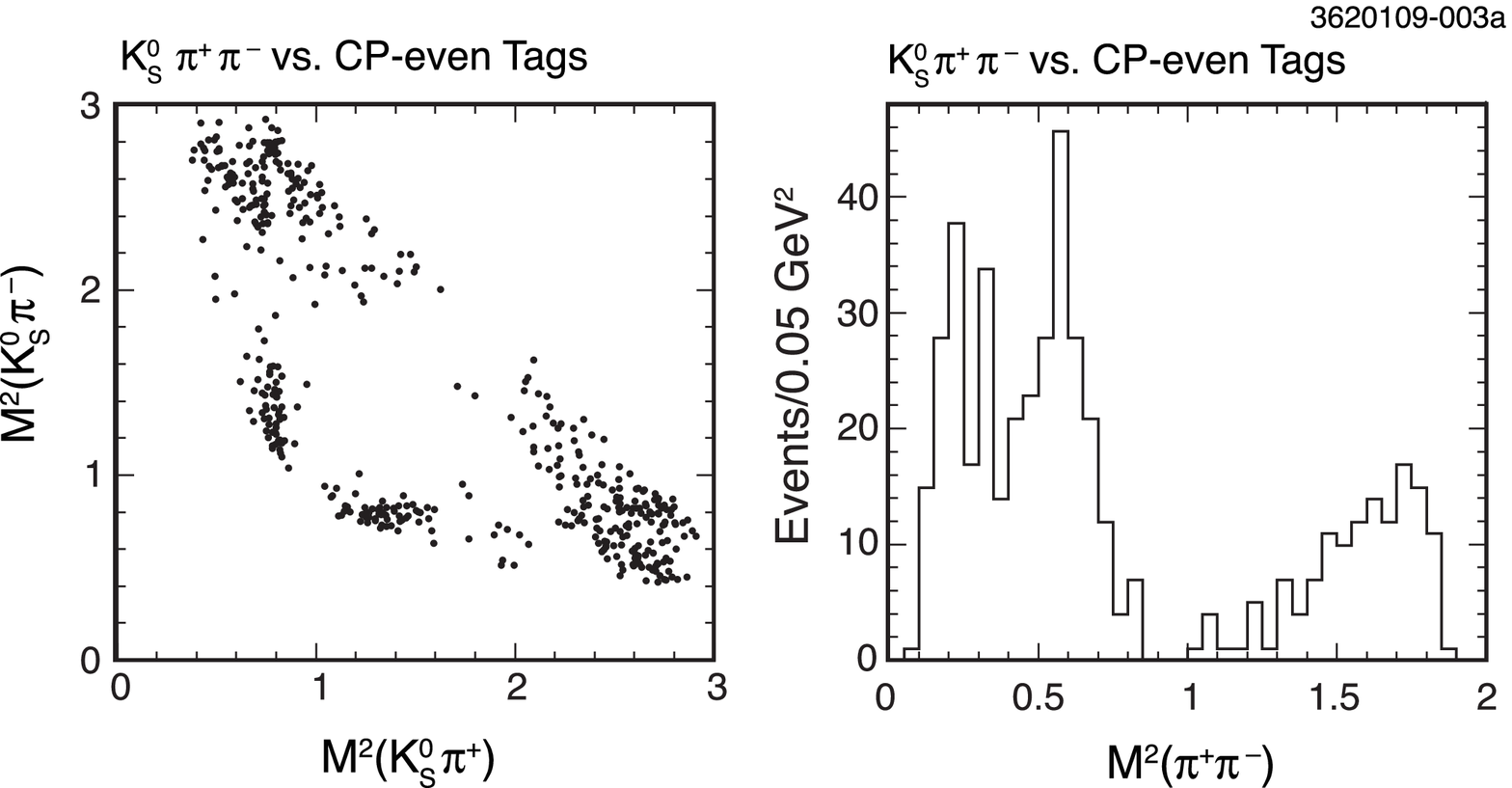}
\includegraphics*[width=0.48\textwidth]{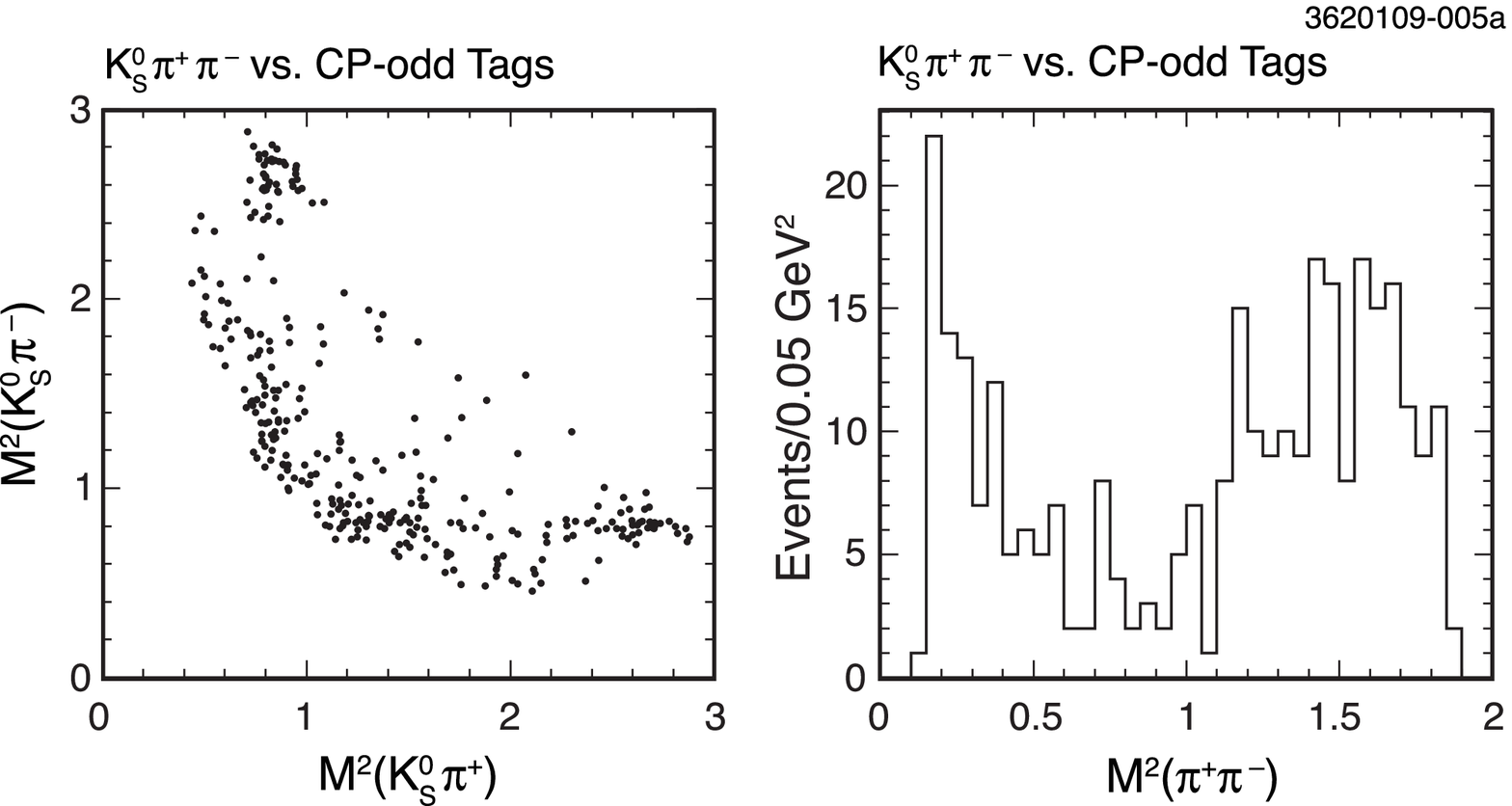}
\caption{CLEO-c Dalitz plots and projections for CP-tagged $K^0_S \pi^+\pi^-$ events.}
\label{fig:kspp_cp}
\end{figure}

\begin{figure}[htb]
\centering
\includegraphics*[width=0.48\textwidth]{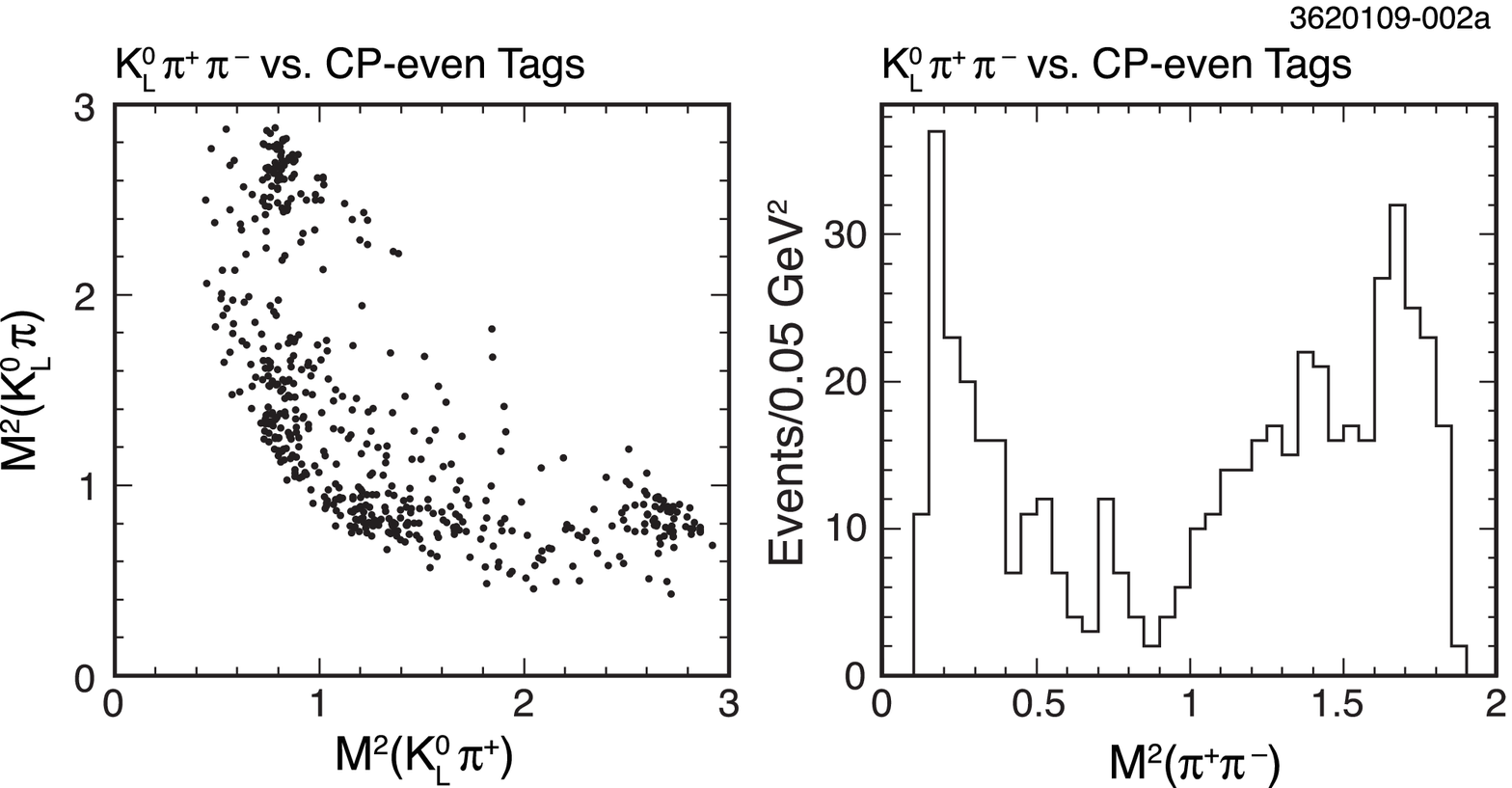}
\includegraphics*[width=0.48\textwidth]{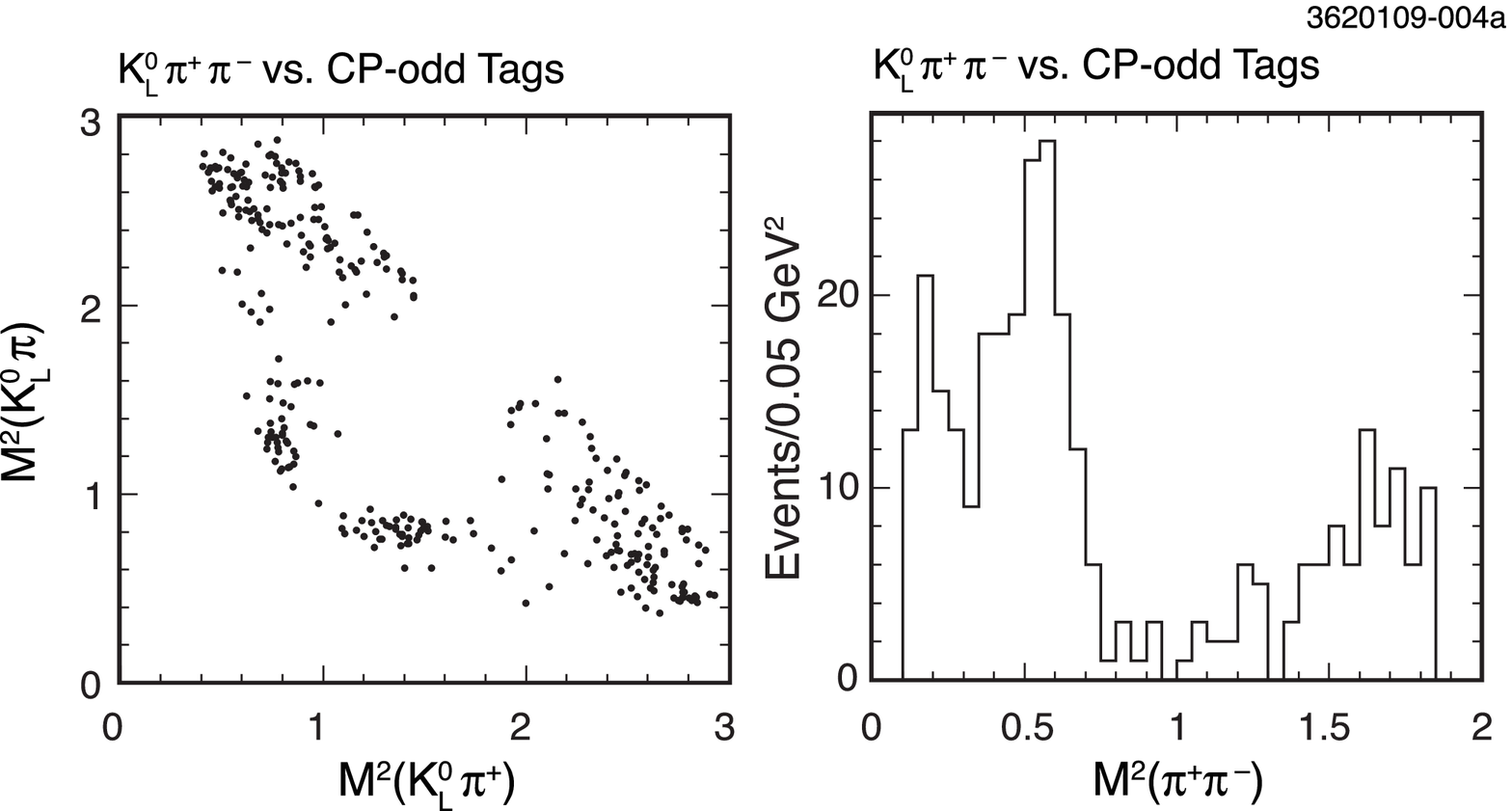}
\caption{CLEO-c Dalitz plots and projections for CP-tagged $K^0_L \pi^+\pi^-$ events.}
\label{fig:klpp_cp}
\end{figure}

An inspection of the Dalitz plots and projections for the 
CP-tagged $K^0_S \pi^+\pi^-$ samples, as presented in Fig.~\ref{fig:kspp_cp},
reveals clear differences.   For example, the sample with the CP-even tags
has in the $m^2(\pi^+\pi^-)$ projection a clear $\rho^0$ peak associated
with the CP-odd $K^0_S \rho$ decays.  This feature is absent from the sample
with the CP-odd tags.  The corresponding $K^0_L \pi^+\pi^-$ distributions
are shown in Fig.~\ref{fig:klpp_cp}. Observe that, as expected, the $K^0_L \pi^+\pi^-$
with CP-even(-odd) tag distributions resemble those of the $K^0_S \pi^-\pi^+$ with CP-odd(-even) tags.

In the analysis the parameters $c_i$ and $s_i$ are determined from measuring
the event yield, after background subtraction and efficiency
correction, in each bin of the Dalitz plot for the CP-tagged 
and the $K^0_S \pi^+\pi^-$ vs $K^0 \pi^+\pi^-$ samples.   The number of events
in bin $i$ of a CP-tagged $K^0_S \pi^+\pi^-$ Dalitz plot is
\begin{equation}
M_i^\pm = h_{CP\pm} \left( K_i \pm 2c_i \sqrt{K_i K_{-i}} + K_{-i} \right)
\label{eq:ksppvscp}
\end{equation}
where $h_{CP\pm}$ is a normalisation factor which can be determined from the
number of single flavour-tagged signal decays and single CP-decays. 
For a $K^0_S \pi^+\pi^-$ vs $K^0_S \pi^+\pi^-$ sample, the number of
events with entries in bin $i$ of the first plot and bin $j$ of the second 
plot is given by
\begin{eqnarray}
M_{ij} & = & h_{corr} ( K_i K_{-j} \, + \, K_{-i}K_{j} \, -  \nonumber \\
& &  2 \sqrt {K_i K_{-j} K_{-i}K_j} (c_i c_j  \, + \,s_i s_j) )
\label{eq:ksppvskspp}
\end{eqnarray}
where $h_{corr}$ is another normalisation factor.

Events including $D^0 \to K^0_L \pi^+\pi^-$ decays make a significant contribution
to the overall sensitivity of the analysis.  Superficially, CP-even (-odd)
$K^0_L \pi^+\pi^-$ events can be treated as CP-odd (-even) $K^0_S \pi^+\pi^-$
events.  The expression for the number of CP-tagged $K^0_L \pi^+\pi^-$
events in bin $i$ is given by
\begin{equation}
M_i^\pm = h_{CP\pm} \left( K_i \mp 2{c_i}' \sqrt{K_i K_{-i}} + K_{-i} \right)
\label{eq:klppvscp}
\end{equation}
and the corresponding expression to Eqn.~\ref{eq:ksppvskspp} for 
$K^0_S \pi^+\pi^-$ vs $K^0_L \pi^+\pi^-$ decays is
\begin{eqnarray}
M_{ij} & = & h_{corr} ( K_i K_{-j} \, + \, K_{-i}K_{j} \, +  \nonumber \\
& &  2 \sqrt {K_i K_{-j} K_{-i}K_j} (c_i {c_j}' \, + \,s_i {s_j}') ).
\label{eq:ksppvsklpp}
\end{eqnarray}
As well as the sign-flips with respect to the earlier expressions, note 
that the cosine and sine of the binned strong phases for the 
$D \to K^0_L \pi^+\pi^-$ decays are denoted with the primed quantities
${c_i}'$ and ${s_i}'$.  These are not expected to be quite identical to $c_i$
and $s_i$, as can be seen by writing
\begin{eqnarray}
A(D^0 \to K^0_S \pi^+\pi^-) &=& \frac{1}{\sqrt{2}}[A(D^0 \to \bar{K^0} \pi^+\pi^-) \nonumber  \\
& + & A(D^0 \to K^0 \pi^+\pi^-)] \nonumber
\end{eqnarray}
and
\begin{eqnarray}
A(D^0 \to K^0_L \pi^+\pi^-) &=& \frac{1}{\sqrt{2}}[A(D^0 \to \bar{K^0} \pi^+\pi^-) \nonumber \\
& - & A(D^0 \to K^0 \pi^+\pi^-)], \nonumber
\end{eqnarray}
from which it follows
\begin{eqnarray}
A(D^0 \to K^0_L \pi^+\pi^-) &=& A(D^0 \to K^0_S \pi^+\pi^-)  \nonumber \\
& - & \sqrt{2} A(D^0 \to K^0 \pi^+\pi^-). \nonumber
\end{eqnarray}
Thus in relating $D^0 \to K^0_L \pi^+\pi^-$ to
$D^0 \to K^0_S \pi^+\pi^-$ the set of doubly Cabibbo suppressed (DCS) amplitudes $A(D^0 \to K^0 \pi^+\pi^-)$ appear as a 
correction term, with a minus sign.  

If, for the purposes of illustration, we consider only
intermediate resonances of the sort $K^{*\pm}$ and $\rho^0$ then it is easy to show
\begin{equation}
A(D^0 \to K^0_S \pi^+\pi^-) = \alpha K^{*-} \pi^+ \, + \, \beta K^{*+}\pi^- \, + \, \chi \rho^0 K^0_S
\end{equation}
and
\begin{equation}
A(D^0 \to K^0_L \pi^+\pi^-) = \alpha K^{*-} \pi^+ \, - \, \beta K^{*+}\pi^- \, + \, \chi' \rho^0 K^0_L,
\end{equation}
where $\alpha$, $\beta$, $\chi$ and $\chi'$ are coefficients. 
Thus in going from $K^0_S \pi^+\pi^-$ to $K^0_L \pi^+\pi^-$ the $K^{*+}\pi^-$ term changes sign, and
the $\rho^0$ term enters with a different factor, which is caused by the sign-flip in the DCS contribution
to this amplitude.  Therefore is is expected that ${c_i}' \ne c_i$ and ${s_i}' \ne s_i$, although the difference
between the two sets of parameters is predicted to be small.

In the analysis ${c_i},{s_i},{c_i}'$ and ${s_i}'$ are extracted simultaneously,  but the 
allowed differences between the unprimed and primed quantities are constrained in the fit.  
The expected differences on which this constraint is based are calculated from the
flavour-tagged models of the $K^0_S \pi^+\pi^-$ decay, which give the coefficients 
$\alpha$, $\beta$ and $\chi$.  The coefficient $\chi'$ is related to $\chi$ by a DCS
correction, the phase and magnitude of which is allowed to vary in a conservative range as a systematic
in the study.   Additional contributions to the systematic uncertainty come from
using different models to give the values of $\alpha$, $\beta$ and $\chi$.  
The results of these studies indicate that the residual model dependence is small
compared with the statistical uncertainty of the analysis.

\subsection{Results and impact on the $\gamma$ measurement}

\begin{figure}[htb]
\centering
\includegraphics*[width=0.38\textwidth]{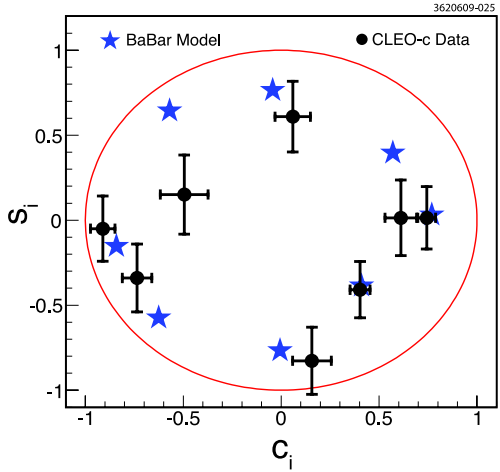}
\caption{CLEO-c results~\cite{CLEOCCISI} and model predictions~\cite{BABARDALITZ} for $c_i$ and $s_i$.}
\label{fig:kpp_results}
\end{figure}

The results of the CLEO-c analysis for $c_i$ and $s_i$ 
are shown in Fig.~\ref{fig:kpp_results},
together with the expectations from the BABAR model~\cite{BABARDALITZ}.
The measurement errors are the sum in quadrature of statistical
uncertainties, uncertainties arising from the reconstruction (such
as that arising from the momentum resolution), and the uncertainties arising from
the residual model dependence in the $K^0_S\pi^+\pi^-$--$K^0_L\pi^+\pi^-$
constraint.  The statistical uncertainties are dominant.  The 
$c_i$ measurements are more precise than those from $s_i$,
as they benefit both from the CP-tags and the 
$K^0 \pi^+\pi^-$ vs $K^0 \pi^+\pi^-$ events,
whereas sensitivity to $s_i$ comes solely from the latter category.
The measurements are compatible with the model predictions.

When these results for $c_i$ and $s_i$ will be used as input to the $\gamma$ determination
using $B \to D(K_S^0 \pi^+\pi^-)K$ decays, the uncertainties
on the parameter values will induce a corresponding error on $\gamma$.
This uncertainty has been estimated to be around $2^\circ$, which is
much smaller than both the present BABAR assigned model uncertainty of $7^\circ$
and the expected statistical uncertainty of $5.5^\circ$ at LHCb with $10~{\rm fb^{-1}}$.
The loss in statistical precision from the binned method, compared with
a binned model-dependent approach, is a relative 20\%.  It is being
investigated whether an alternative choice of binning could reduce this loss. 
With the present binning the sensitivity to $\gamma$ will surpass  
that of the unbinned approach with less than $2~{\rm fb^{-1}}$ of LHCb data.
\vspace*{-0.2cm}
\subsection{Extending to $D \to K^0_S K^+K^-$}
As has been demonstrated by BABAR~\cite{BABARDALITZ}, $B \to D(K^0_S K^+K^-)K$
decays can be used to measure $\gamma$ using a model dependent unbinned fit with
a method entirely analogous to that used for $B \to D(K^0_S \pi^+\pi^-)K$.
In order to allow a model independent exploitation of this mode, CLEO-c has
embarked upon a measurement of the corresponding $c_i$ and $s_i$ parameters
in $D \to K^0_S K^+K^-$.  This determination will exploit around 550 quantum-correlated
double tags, including $K^0 K^+ K^-$ vs $K^0 \pi^+ \pi^-$ events that
can contribute to the analysis thanks to the knowledge of the  
$c_i$ and $s_i$ values for $D \to K^0_S \pi^+\pi^-$.

\begin{figure}[htb]
\centering
\includegraphics*[width=0.24\textwidth]{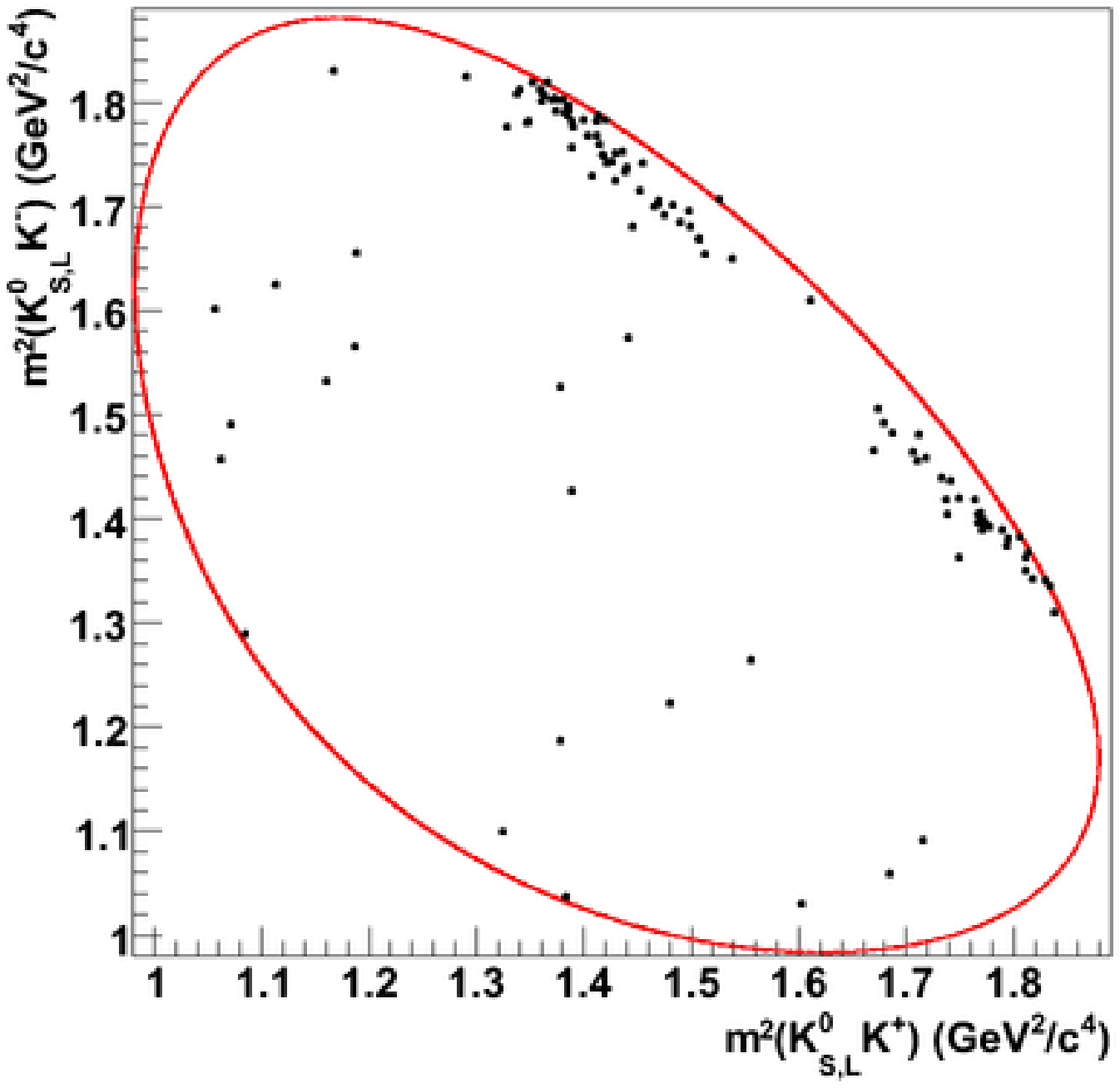}
\includegraphics*[width=0.24\textwidth]{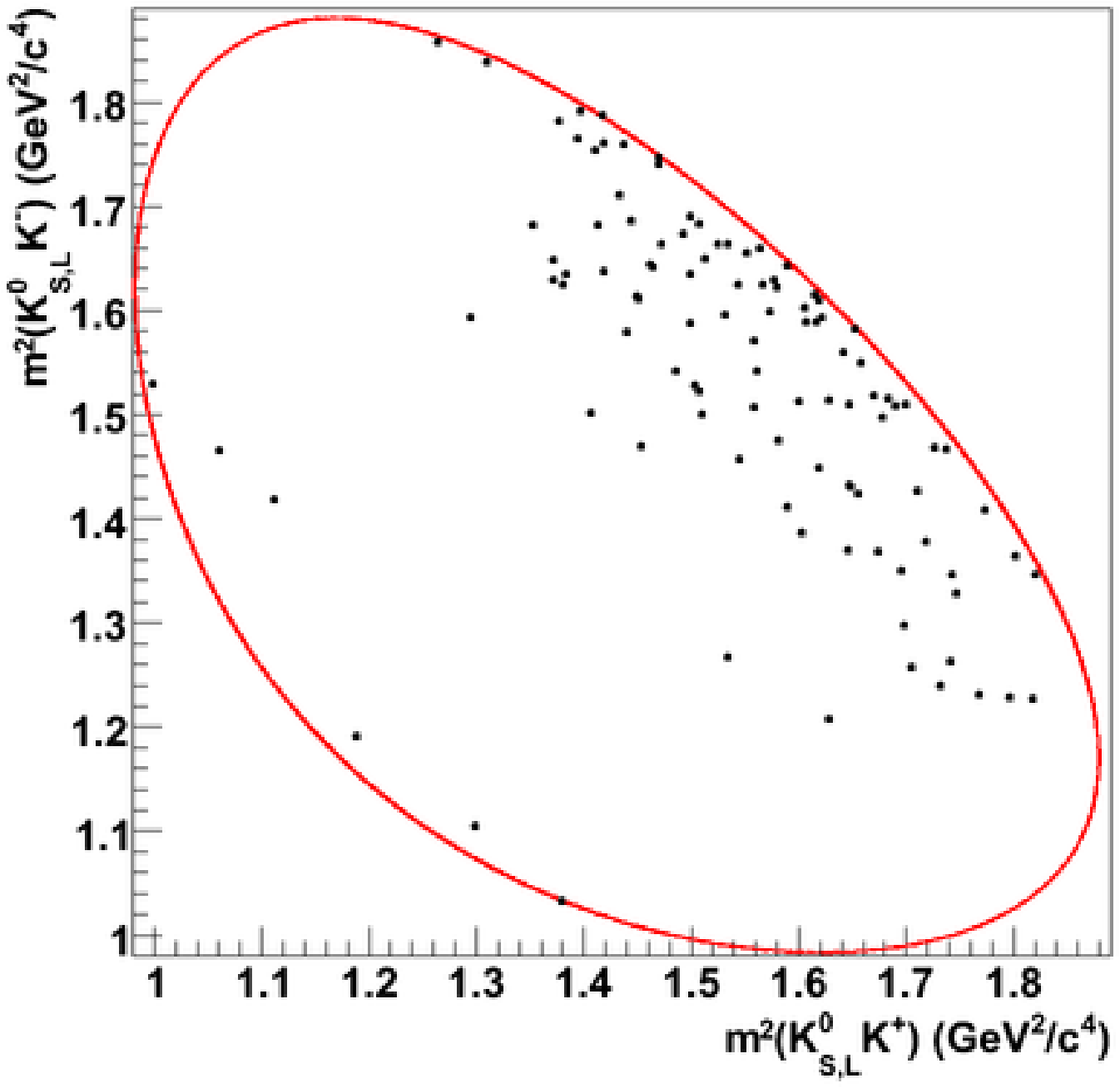}
\includegraphics*[width=0.24\textwidth]{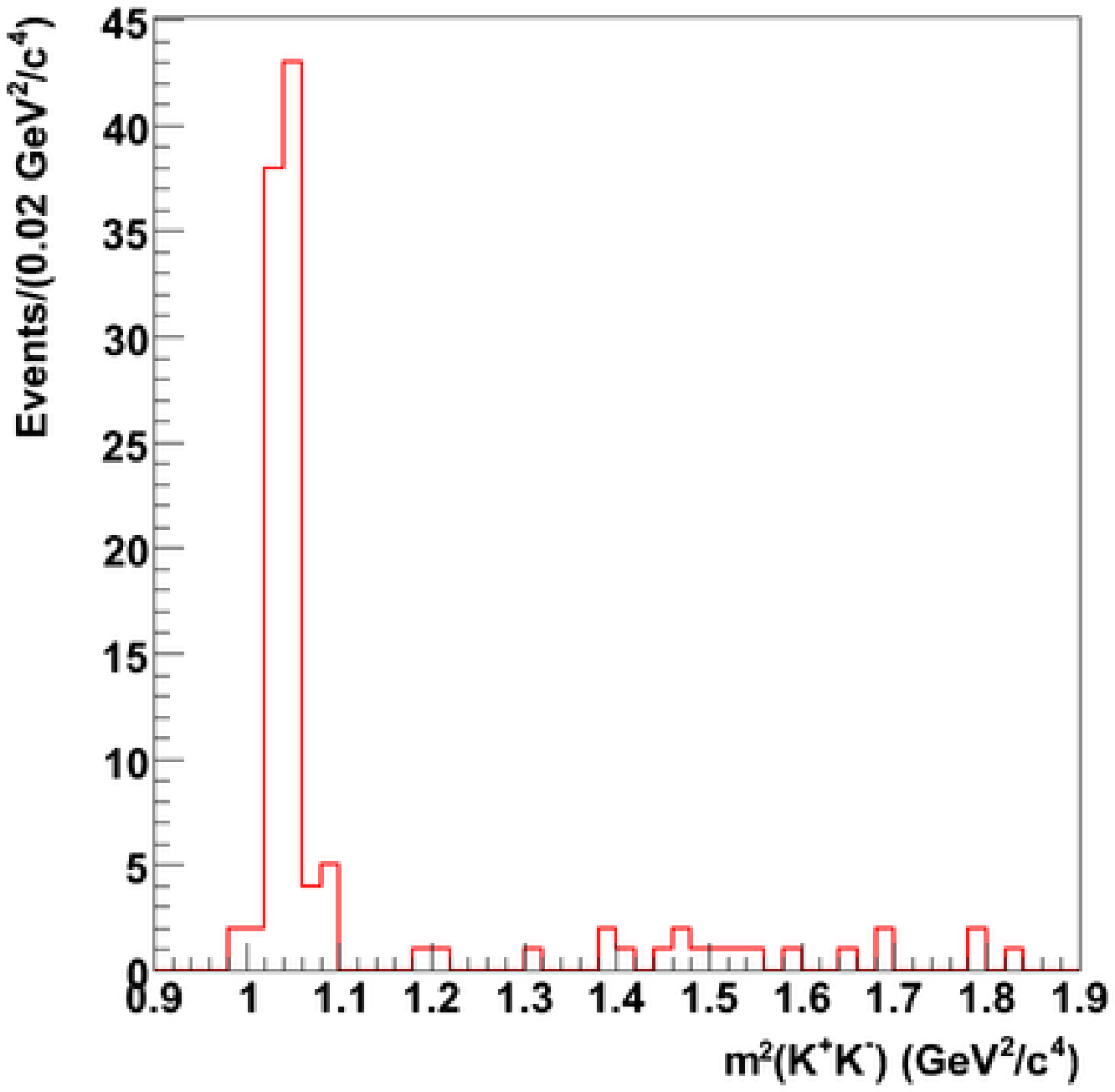}
\includegraphics*[width=0.24\textwidth]{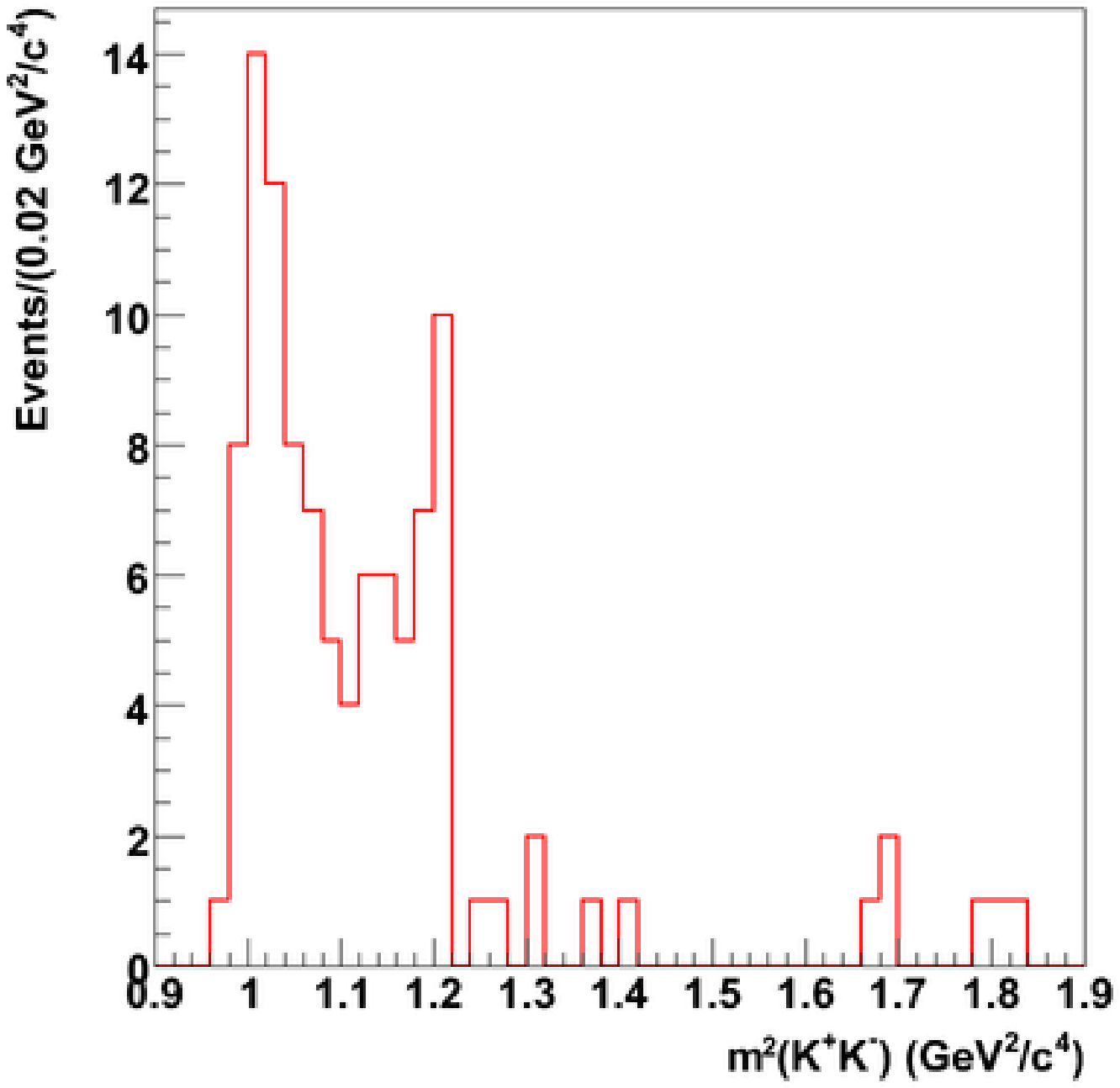}
\caption{CLEO-c $D \to K^0 K^+K^-$ Dalitz plots and projections.
Left: $K^0_{S(L)} K^+ K^-$ with CP-even(odd) tags.
Right: $K^0_{S(L)} K^+ K^-$ with CP-odd(even) tags.}
\label{fig:kskk}
\end{figure}

Figure~\ref{fig:kskk} shows the Dalitz plots and $m^2(K^+K^-)$ projections
for CP-tagged $K^0_S K^+K^-$ and $K^0_L K^+K^-$ events.  Observe that
the $\phi$ peak associated with  $K^0_{S (L)} \phi $ decays is only prominent for
the CP-even(odd) tags.

Preliminary results on the measurement of the strong phase difference in
$D \to K^0_S K^+K^-$ events can be found in~\cite{STEFANIA}.

\vspace*{-0.6cm}
\section{Quantum correlated studies of \\ $D \to K (n) \pi$ decays}

\subsection{The ADS strategy for measuring $\gamma$}
A powerful subset of the $B\to DK$ family of methods to measure $\gamma$
is the so-called `ADS' approach~\cite{ADS}.  Here the $D$-decay mode
that is reconstructed is one which involves a charged kaon and one or
more pions.  The simplest example is $D \to K^\pm \pi^\mp$, which
is here taken as an example.  Depending
on the charge of the $B$-meson and the kaon from the $D$-decay, there
are four possible final states.  The partial widths into each state
depend on the physics parameters of interest. Two of these final states
have particular sensitivity:
\begin{eqnarray}
\Gamma (B^\mp \to (K^\pm\pi^\mp)_D K^\mp)  & \propto &  \nonumber 
\end{eqnarray}
\vspace*{-0.8cm}
\begin{eqnarray}
r_B^2 + (r_D^{K\pi})^2  + 2 r_B r_D^{K\pi}  \cos(\delta_B + \delta^{K\pi}_D \mp \gamma). &&
\label{eq:adskpi}
\end{eqnarray}
Since $r_B$ and the magnitude of the ratio between the doubly Cabibbo suppressed  and the favoured
$D$ decay amplitudes, $r_D^{K\pi}$, are of similar size, the interference
term that involves $\gamma$ appears at first order.  The consequence is that a large 
asymmetry may exist between the number of events found in each final state. Measuring
this asymmetry, and combining with observables in other $B \to DK$ modes, allows
$\gamma$ to be determined.   This extraction however benefits from external
constraints on the $D$-decay parameters $r_D^{K\pi}$  and $\delta_D^{K\pi}$.
The former of these is well known, essentially from the ratio of the
suppressed and favoured branching ratios.  Knowledge of $\delta_D^{K\pi}$ comes 
both from $\psi(3770)$ decays and from the ensemble of $D$-meson mixing measurements,
as is discussed below.

The form of the ADS decay rates, given in expression~\ref{eq:adskpi}, takes on
a different form for multibody decays such as $D \to K^\pm \pi^\mp \pi^+\pi^-$.
In this case there are many intermediate resonances (eg. $K^{*0}\rho^0$, $K^\pm a_1^\mp$), 
which in general will contribute with different strong phases.  The two rates of 
interest are then as follows:
\begin{eqnarray}
\Gamma (B^\mp \to (K^\pm\pi^\mp\pi^-\pi^+)_D K^\mp)  & \propto &  \nonumber 
\end{eqnarray}
\vspace*{-0.8cm}
\begin{eqnarray}
r_B^2 + (r_D^{K3\pi})^2  + 2 r_B r_D^{K3\pi} R_{K3\pi} \cos(\delta_B + \delta^{K3\pi}_D \mp \gamma). &&
\label{eq:adscoherence}
\end{eqnarray}
The parameter $R_{K3\pi}$ is termed the coherence factor, and can take any value between $0$ and $1$,
where the latter limit corresponds to the case when all resonances contribute in phase and the 
channel behaves as a two-body decay.  The parameter $\delta^{K3\pi}_D$ is now 
the strong-phase difference averaged over all Dalitz space.  Precise definitions of the quantities
can be found in~\cite{COHERENCE}.  Analogous parameters exist for other decays, for example
$R_{K\pi\pi^0}$ and $\delta^{K\pi\pi^0}_D$ in the case of $D \to K^\pm \pi^\mp \pi^0$.

\subsection{Analysis of the mode $D \to K^\pm \pi^\mp$}

The strong-phase difference $\delta_D^{K\pi}$ between $D^0$ and $\bar{D^0}$ decays to $K^+\pi^-$ has
been determined by CLEO-c using 281~$\rm {pb^{-1}}$ of $\psi(3770)$ data~\cite{TQCA}.
This result is the least recent of those reported in this review, and so only a very brief
summary of the analysis is given here.
 
If one neglects mixing then the rate, $F^{K\pi}_{CP \pm}$, of 
CP-tagged $D \to K^\pm \pi^\mp$ events is as follows:
\begin{equation}
F^{K\pi}_{CP \pm} \approx {\mathcal{B}}_{CP \pm} {\mathcal{B}_{K\pi}}(1 + (r_D^{K\pi})^2 \pm 2 r_D^{K\pi}\cos\delta_D^{K\pi}).
\label{eq:tqca}
\end{equation}
where ${\mathcal{B}}_{CP \pm}$ and ${\mathcal{B}_{K\pi}}$ are the branching ratios of the CP-tag
and the Cabibbo favoured signal decay, respectively.  The analysis reported in~\cite{TQCA} uses a range of 
CP-tags, other double-tagged events, and single tags both to extract $\delta_D^{K\pi}$ and to gain
sensitivity to the mixing parameters $x$ and $y$, which modify the result for $F^{K\pi}_{CP \pm}$ and
the rates for other event types.

\begin{figure}[htb]
\centering
\includegraphics*[width=0.38\textwidth]{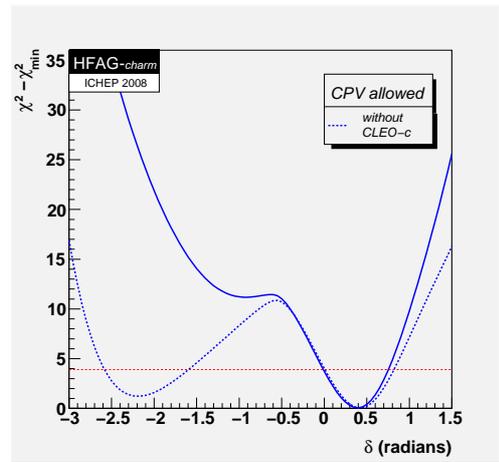}
\caption{CLEO-c impact on the knowledge of $\delta_D^{K\pi}$~\cite{HFAG}.
The dotted line is obtained from a global fit to $D$-mixing measurements
which do not use $\psi(3770)$ data and the solid line from a fit which in addition includes 
the result of~\cite{TQCA}.}
\label{fig:hfag_tqca}
\end{figure}

When a fit is performed which imposes no external constraints on the mixing parameters, a result
for the strong phase difference of $\cos \delta_D^{K\pi} = 1.03^{+0.31}_{-0.17} \pm 0.06$ 
is obtained~\footnote{It is worth remarking that the convention used to describe the effect of a
CP operation on the $D^0$ meson has non-trivial consequences in the definition of phase differences.
In particular, the convention assumed in most charm mixing analyses, and implicit in the results presented here, is
offset by $\pi$ from that assumed in most $B \to DK$ studies.  Thus the central value that should be used 
in both expressions~\ref{eq:adskpi} and~\ref{eq:adscoherence} is {\it not} $\delta_D^{K\pi} \approx 0.4$, as suggested
by Fig.~\ref{fig:hfag_tqca}, but $\delta_D^{K\pi} \approx 0.4 - \pi$.  }.   In fact this measurement is less precise than that of 
the indirect determination
which can be obtained from a global fit to all the $D$-mixing results~\cite{HFAG}. As Fig.~\ref{fig:hfag_tqca} 
makes clear, however, the inclusion
of the CLEO-c result brings important extra information,  as it resolves a two-fold
ambiguity which would otherwise exist in our knowledge of $\delta^{K\pi}_D$.

This analysis is now being updated with the full CLEO-c dataset and improved
analysis methods~\cite{NAIK}.

\subsection{Analysis of the modes 
$D \to K^\pm \pi^\mp \pi^+\pi^-$ and $D \to K^\pm \pi^\mp \pi^0$}

The coherence factors and the average strong phase differences have been
determined by CLEO-c for the modes $D \to K^\pm \pi^\mp \pi^+\pi^-$ and $D \to K\pm \pi^\mp \pi^0$
using 818~${\rm fb^{-1}}$ of $\psi(3770)$ data~\cite{CLEOCCOHERENCE}.  The analysis is based on
the method proposed in~\cite{COHERENCE}.  Each signal mode is reconstructed alongside various 
categories of tags.  These include: CP-eigenstates; the signal mode itself in the case
where the two kaons in the event are of identical sign (giving so-called `like-sign' events); 
the other signal mode under consideration, again in the like-sign configuration;
and like-sign $K^\pm \pi^\mp$ decays.  The sensitivity of each event class to the coherence
factors and strong phase differences is indicated in Tab.~\ref{tab:coherence_sens},
although Ref.~\cite{COHERENCE} should be consulted to obtain the complete expressions.

\begin{table}
\begin{center}
\caption{Dependence of double-tag rates in the coherence factor analysis.}
\label{tab:coherence_sens}
\begin{tabular}{ll} \hline \hline
Double-tag   & Sensitive to \\ \hline
$K^\pm \pi^\mp \pi^+ \pi^-$ vs $K^\pm \pi^\mp \pi^+\pi^-$  & $(R_{K3\pi})^2$ \\
$K^\pm \pi^\mp \pi^0 $ vs $K^\pm \pi^\mp \pi^0$            & $(R_{K\pi\pi^0})^2$ \\
$K^\pm \pi^\mp \pi^+ \pi^-$ vs CP                          & $R_{K3\pi} \cos(\delta_D^{K3\pi})$ \\
$K^\pm \pi^\mp \pi^0$ vs CP                                & $R_{K\pi\pi^0} \cos(\delta_D^{K\pi\pi^0})$ \\
$K^\pm \pi^\mp \pi^+\pi^-$ vs $K^\pm \pi^\mp$              & $R_{K3\pi} \cos(\delta_D^{K3\pi} - \delta_D^{K\pi})$ \\
$K^\pm \pi^\mp \pi^0$ vs $K^\pm \pi^\mp$                   & $R_{K\pi\pi^0} \cos(\delta_D^{K\pi\pi^0} - \delta_D^{K\pi})$ \\
$K^\pm \pi^\mp \pi^+\pi^-$ vs $K^\pm \pi^\mp \pi^0$         & $R_{K3\pi} R_{K\pi\pi^0} \cos(\delta_D^{K3\pi} - 
\delta_D^{K\pi\pi^0})$
\end{tabular}
\end{center}
\end{table}

The analysis uses 10 types of CP-tags.  The event yield for each category of event,
after background subtraction, is listed in Tab.~\ref{tab:coherenceyield}.
Additional double-tags, not detailed here, include `unlike-sign' events,
and $K^\pm \pi^\mp$ vs CP-eigenstate events, both needed for normalisation purposes. 

\begin{table}
\vspace*{-0.4cm}
\caption{CLEO-c double-tag background subtracted 
yields in the coherence factor analysis analysis.}\label{tab:coherenceyield}
\begin{center}
\vspace*{-0.2cm}
\begin{tabular}{lcc}\\ \hline \hline
Tag Mode   & $K^\pm \pi^\mp \pi^+\pi^-$ yield & $K^\pm \pi^\mp \pi^0$ yield \\ \hline
\multicolumn{3}{c}{CP-even tags} \\ \hline
$K^+K^-$              & 536 &  764 \\
$\pi^+\pi^-$          & 246 &  336 \\
$K^0_S \pi^0\pi^0$    & 283 &  406 \\
$K^0_L\pi^0$          & 695 & 1234 \\ 
$K^0_L\omega$         & 296 &  449 \\ \hline
\multicolumn{3}{c}{CP-odd tags} \\ \hline
$K^0_S \pi^0$         & 705 & 891 \\
$K^0_S \omega$        & 319 & 389 \\
$K^0_S \phi$          &  53 &  91 \\ 
$K^0_S \eta$          & 164 & 152 \\  
$K^0_S \eta'$         &  36 &  61 \\ \hline
\multicolumn{3}{c}{Other tags} \\ \hline
$K^\pm \pi^\mp \pi^+\pi^-$ & 29 & 64 \\
$K^\pm \pi^\mp \pi^0$      & see row above & 13 \\
$K^\pm \pm^\mp$            & 36 &  7 \\
\hline
\end{tabular}
\vspace*{-0.3cm}
\end{center}
\end{table}

\begin{figure}[htb]
\centering
\includegraphics*[width=0.40\textwidth]{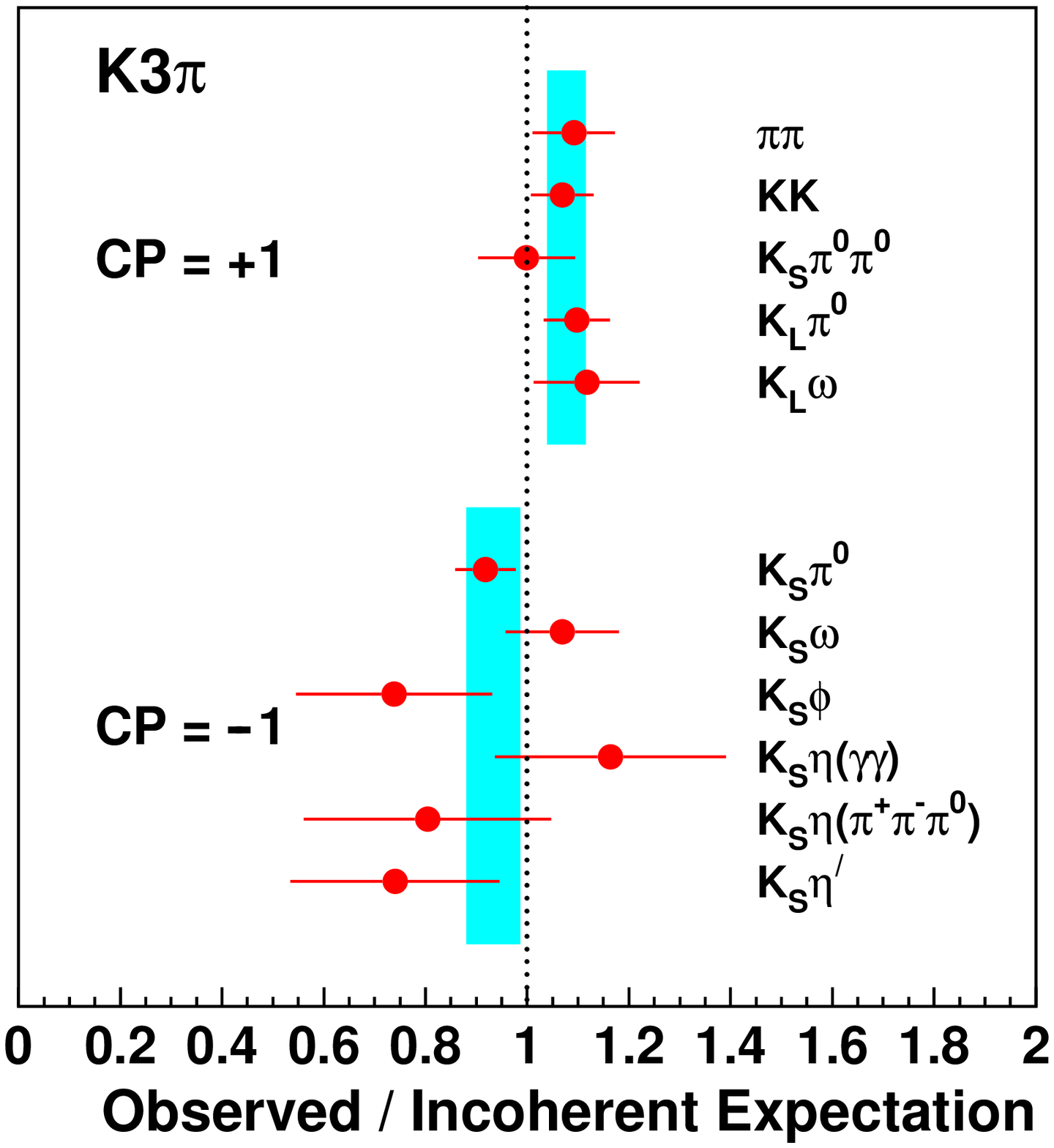}
\includegraphics*[width=0.40\textwidth]{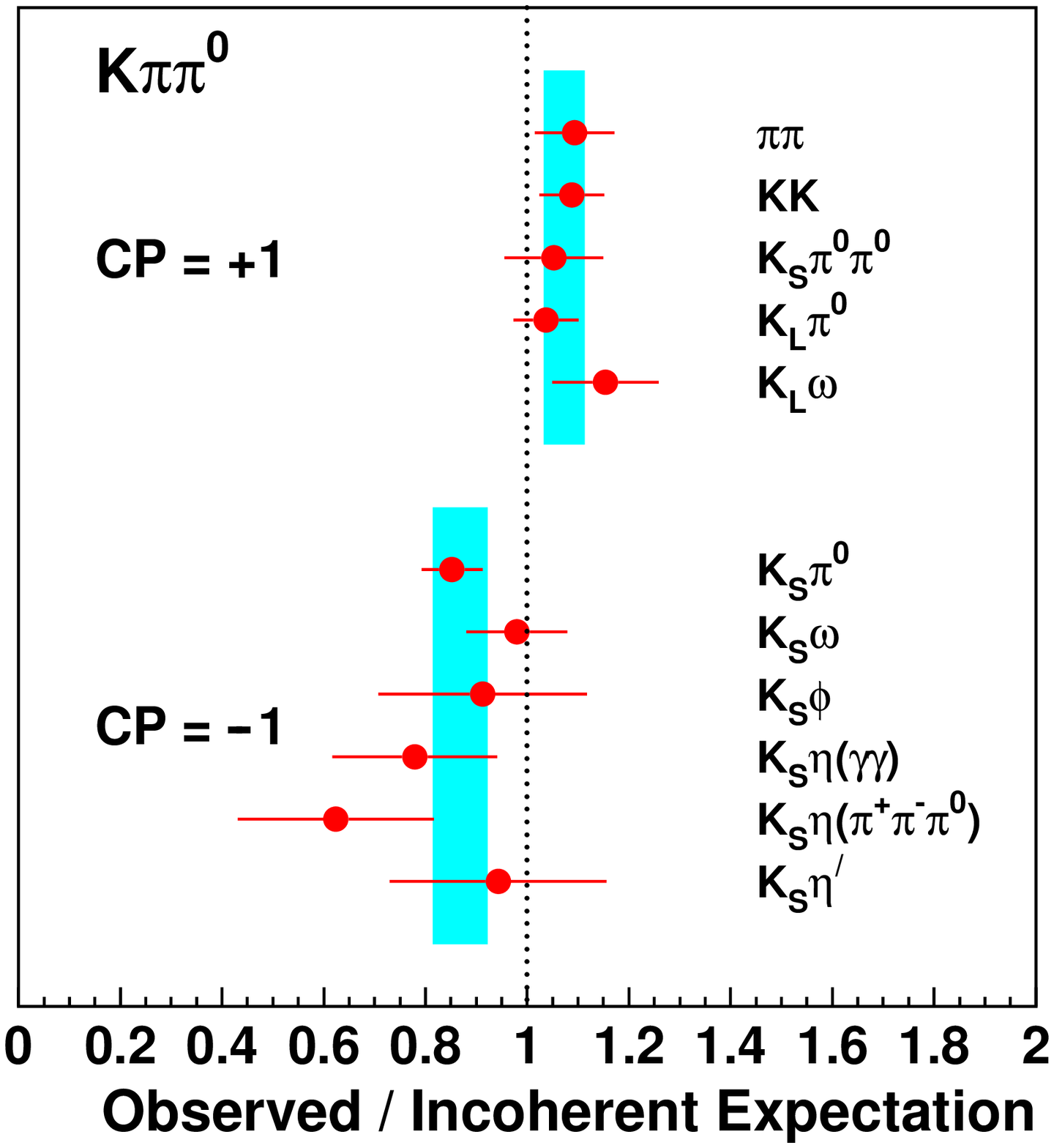}
\caption{CLEO-c $\rho_{\rm CP}$ parameters for the $D \to K^\pm \pi^\mp \pi^+\pi^-$ (top) 
and $D \to K^\pm \pi^\mp \pi^0$ (bottom) analyses, representing the number of 
observed events to the incoherent expectation.  The cyan shaded bands show the
mean result for each CP-eigenvalue.}
\label{fig:coherence_cp}
\end{figure}

The analysis chooses as observables so-called `$\rho$-parameters', which
give the ratio of the number of observed events in each category to
the number expected were the $D-\bar{D}$ pair to decay in an uncorrelated
manner, and/or the coherence parameter to be zero.   
The value of $\rho_{\rm CP}$,
this ratio for the CP-tagged events, is shown for each CP-tag in Fig.~\ref{fig:coherence_cp}.
For a given signal-mode the same behaviour is expected for each CP-eigenvalue,
and behaviour of an opposite sign for CP-odd and CP-even.  The values of the observables
are consistent with these expectations.

\begin{figure}[htb]
\centering
\includegraphics*[width=0.48\textwidth]{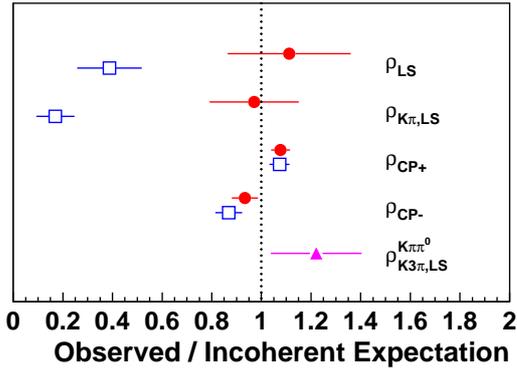}
\caption{CLEO-c results for coherence observables.  Filled red circles indicate 
$D \to K^\pm \pi^\mp \pi^+\pi^-$, open blue squares indicate 
$D \to K^\pm \pi^\mp \pi^0$ and the filled magenta triangle indicates like-sign
$K^\pm \pi^\mp \pi^+\pi^-$ vs $K^\pm \pi^\mp \pi^0$ events.}
\label{fig:coherence_obs}
\end{figure}

In Fig.~\ref{fig:coherence_obs} are shown the results for the nine observables.
These comprise: the mean results per CP-tag per mode ($\rho_{\rm CP+}$ and $\rho_{\rm CP-}$);
the results for the like-sign events for both signal modes ($\rho_{\rm LS}$);
the results for the like-sign $K\pi$ tags ($\rho_{\rm K\pi, LS}$);
and that coming from the like-sign $K^\pm \pi^\mp \pi^+\pi^-$ vs $K^\pm \pi^\mp \pi^0$ events
($\rho^{K\pi\pi^0}_{K3\pi,LS}$).
The error bars include the statistical and systematic uncertainties.
In the case of $\rho_{\rm CP \pm}$ the largest systematic uncertainty is
associated with the normalisation procedure, which is significant alongside
the statistical uncertainty, but is itself statistical in origin.
For the other observables the systematic uncertainties are small.

The expected size and sign of any deviation of the $\rho$ parameters from a value of one depends on the
tag-category.  The general behaviour in Fig.~\ref{fig:coherence_obs}, however, makes clear that there is evidence of 
high coherence in $D \to K^\pm \pi^\mp \pi^0$ decays, but much less so 
for $D \to K^\pm \pi^\mp \pi^+\pi^-$.

\begin{figure}[htb]
\centering
\includegraphics*[width=0.40\textwidth]{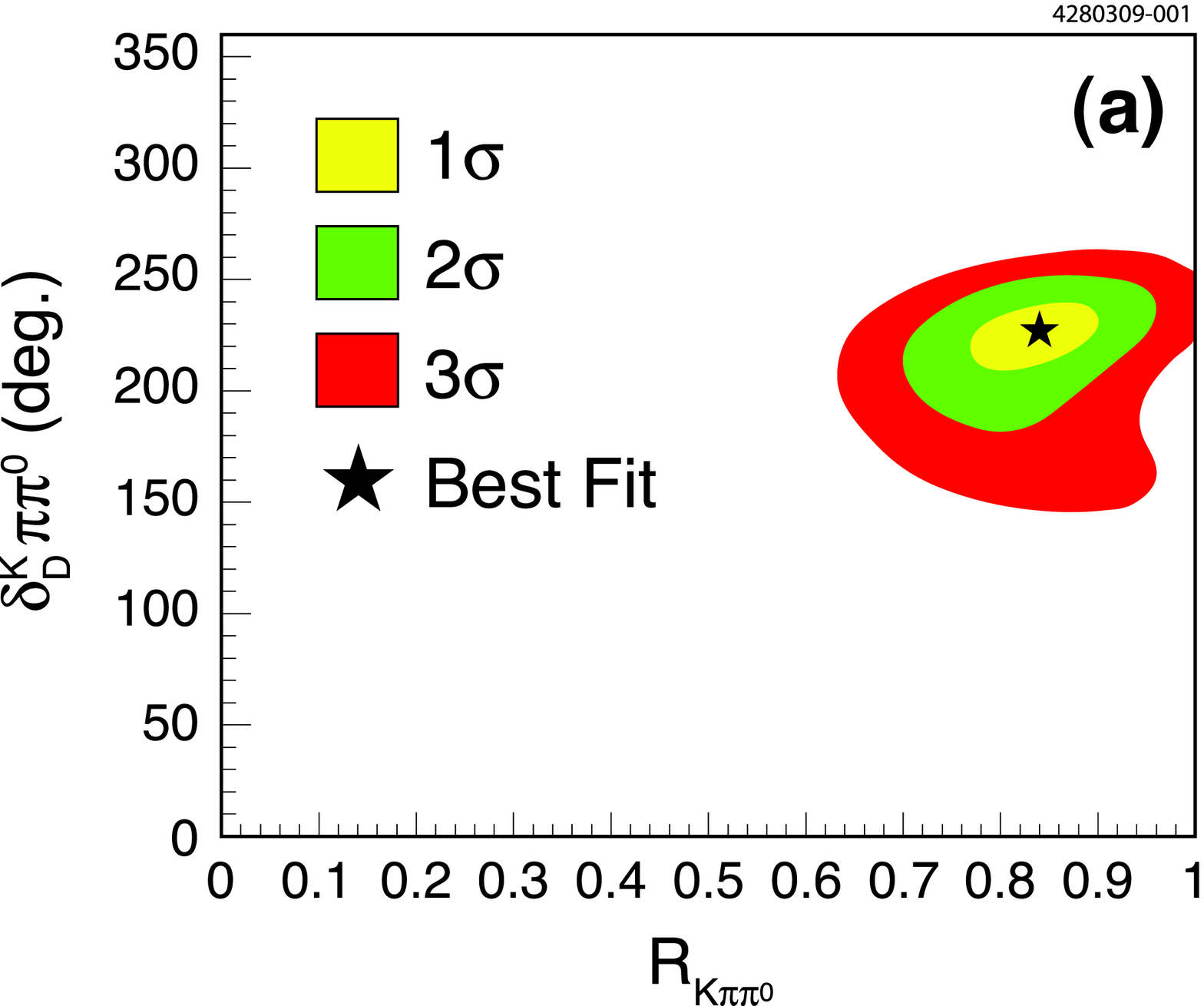}
\includegraphics*[width=0.40\textwidth]{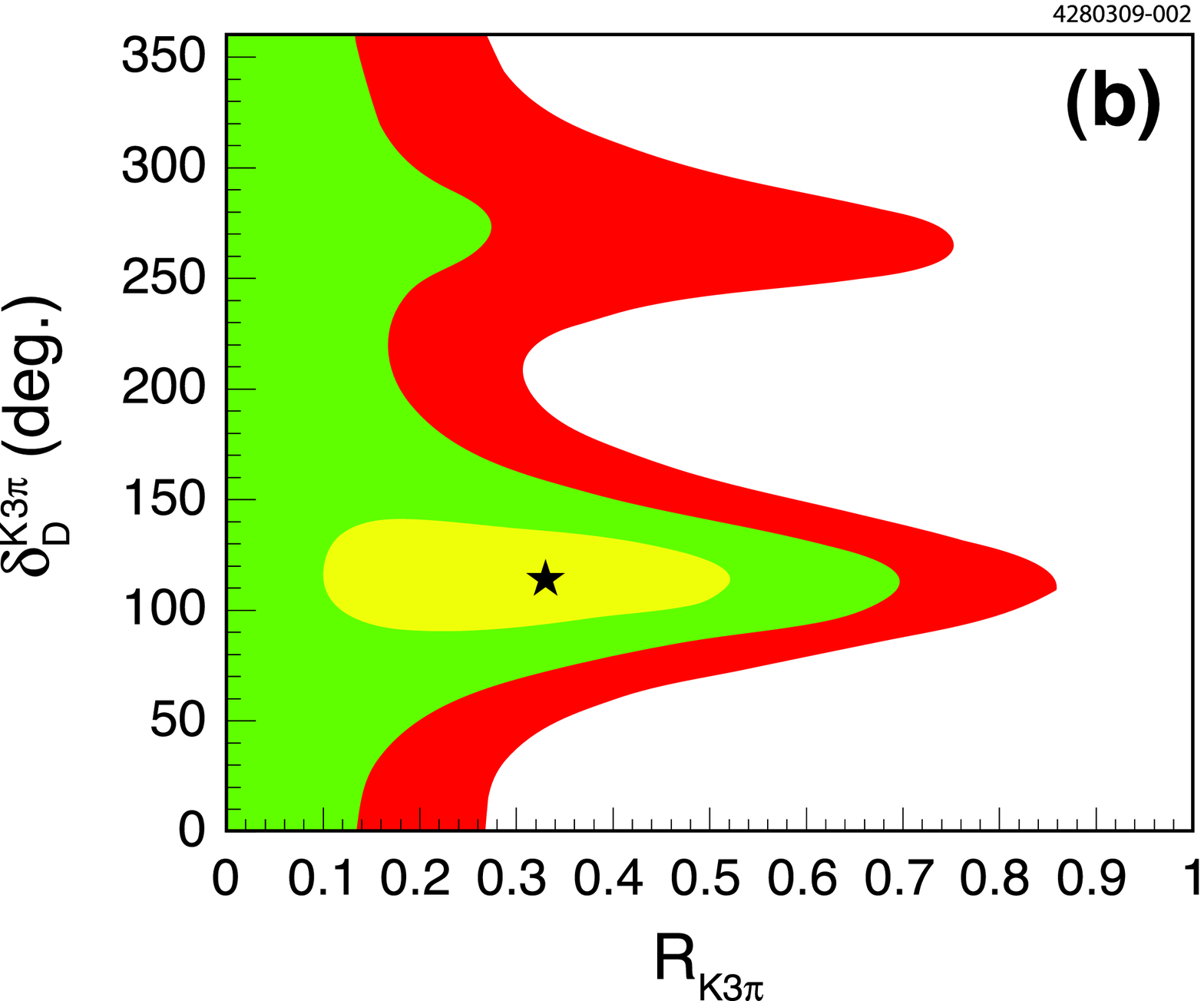}
\caption{CLEO-c results for the coherence factor and mean strong phase difference for
$D \to K^\pm \pi^\mp \pi^0$ (a) and $D \to K^\pm \pi^\mp \pi^+\pi^-$ (b).}
\label{fig:coherence_res}
\end{figure}

Values for the coherence factors, $R_{K3\pi}$ and $R_{K\pi\pi^0}$, and mean strong-phase differences,
$\delta_D^{K3\pi}$ and $\delta_D^{K\pi\pi^0}$, have been obtained by making a $\chi^2$ fit to the 
above observables.  Other free parameters in the fit include $\delta_D^{K\pi}$, the mixing
parameters $x$ and $y$ and the Cabibbo favoured and doubly suppressed branching ratios,
all of which are given a Gaussian constraint to lie close to their world-best measured values.
The best fit values for the coherence factors and strong phases are as follows:
$R_{K3\pi}=0.33^{+0.20}_{-0.23}$, $R_{K\pi\pi^0}=0.84\pm 0.07$, 
$\delta_D^{K3\pi}=(114^{+26}_{-23})^\circ$ and $\delta_D^{K\pi\pi^0} = (227^{+14}_{-17})^\circ$.
The one, two and three sigma contours are shown in Fig.~\ref{fig:coherence_res}.
Thus it is seen that $D \to K^\pm\pi^\mp\pi^0$ is highly coherent, whereas the indications
are that this is not so for $D \to K^\pm\pi^\mp\pi^+ \pi^-$.
Interesting results are also obtained for the auxiliary parameters in the fit,
where in some cases small but significant improvements are found with respect to the
external constraints. For example the fitted value of $\delta_D^{K\pi}$ is $(-151.5^{+9.6}_{-9.5})^\circ$  
to be compared with the applied constraint of $(-157.5^{+10.4}_{-11.0})^\circ$.
This sensitivity arises from the importance of the like-sign $K\pi$ tags in
the analysis.  A relative 10\% improvement is also found in the knowledge of $y$,
with the fit returning a value of $0.81 \pm 0.16 \%$, to be compared with
the applied constraint of $0.76 \pm 0.18 \%$.

\subsection{Impact on $\gamma$ determination}

A study has been made within LHCb to assess the importance of the CLEO-c 
$D \to K (n) \pi$ results on the $\gamma$ measurement~\cite{AKIBA}.
A standalone simulation study has been made in which the precision
on $\gamma$ is determined from a simultaneous analysis of 
$B \to DK$ events using both $K^\pm \pi^\mp$ and 
$K^\pm \pi^\mp \pi^+\pi^-$ as $D$-decay modes. The simulated
events are generated with the $D$-decay parameters set to the central values
of the CLEO-c analysis.  In addition to $\gamma$, the fit also returns $r_B$, the $B$- and $D-$meson
strong phases, and the coherence factors.
The assumed sample size corresponds to
one nominal year (2~${\rm fb^{-1}}$) of data.  The results are compared
between the case where no external knowledge is assumed, and the 
case where the CLEO-c results for $\delta_D^{K\pi}$, $R_{K3\pi}$ and 
$\delta_D^{K3\pi}$ are applied as external constraints in the fit.

The exact results of the study vary with the assumed value of the parameters, but in
general a significant improvement in the precision on $\gamma$ is observed
when the CLEO-c constraints are used,
similar to that which would come about from a doubling of the LHCb dataset.
The impact of the $D \to K^\pm \pi^\mp$ and the $D \to K^\pm \pi^\mp \pi^+\pi^-$
constraints are found to be similar.  At first sight the importance 
of the $D \to K^\pm \pi^\mp \pi^+\pi^-$ events in the analysis is unexpected,
given the low value of the coherence.  This effect can be understood by inspecting 
Eqn.~\ref{eq:adscoherence} and considering the limit when $R_{K3\pi} \to 0$.
In this case the observed decay rate in the suppressed $D \to K^\pm \pi^\mp \pi^+\pi^-$ mode
allows $r_B$ to be determined, which then benefits the $\gamma$ extraction from the 
simultaneous $D \to K^\pm \pi^\mp$ analysis.

The mode $D \to K^\pm \pi^\mp \pi^0$ has not yet been included in the LHCb
ADS analysis, but it is anticipated that here also the CLEO-c constraints
will be helpful in improving the overall $\gamma$ sensitivity.

\section{Summary and prospects}

CLEO-c analyses have been published which determine $D$-decay parameters in
the modes $D \to K^0_S \pi^+\pi^-$, $D \to K^\pm \pi^\mp$, 
$D \to K^\pm \pi^\mp \pi^+\pi^-$ and  $D \to K^\pm \pi^\mp \pi^0$.
All these studies rely on the quantum-correlated nature
of the $D-\bar{D}$ pair in $\psi(3770)$ decays.
The results are found to have significant consequences for the
measurement of $\gamma$ in $B \to DK$ decays,
allowing for both an improvement in overall precision and
the removal of model dependence in the analyses.   Results
are anticipated soon in the mode $D \to K^0_S K^+K^-$.
Work is also underway to extend the $D \to K^\pm \pi^\mp$
analysis to the full 818~${\rm fb^{-1}}$ dataset, and to provide
further results in $D \to K^0_S \pi^+\pi^-$ for alternative
choices of binnings.

There exist other channels which are potentially useful
in the $B \to DK$ analysis and so could benefit
from measurements of their decay properties in quantum
correlated events.  These include $D \to K^+K^-\pi^+\pi^-$,
$D \to K^0_S K^\pm \pi^\mp$ and $D \to K^0_S \pi^+\pi^-\pi^0$

The studies reported here would benefit greatly from the increase
in $\psi(3770)$ statistics which could be collected by the BES-III experiment.
For this reason a significant open charm programme at BES-III is
to be encouraged.

\section*{Acknowledgements}

I am grateful to Jim Libby and other CLEO-c colleagues for useful discussions in preparing this review.


\end{document}